\documentstyle[epsf,fleqn,aps]{revtex}

\def\Section#1{}

\def\beq{\begin{equation}}
\def\eeq{\end{equation}}
\def\bea{\begin{eqnarray}}
\def\eea{\end{eqnarray}}

\begin{document}
\title{Competing interactions in the  $XYZ$ model}
\author{M.\ Arlego$^1$, D.C.\ Cabra$^{2,\#}$, J.E.\ Drut$^1$ and M.D.\
Grynberg$^1$}
\address{$^1$Departamento de F\'{\i}sica, Universidad Nacional de
La Plata, C.C.\ 67, (1900) La Plata, Argentina.\\
$^{2}$Laboratoire de Physique, Groupe de Physique Th\'eorique
ENS Lyon, 46 All\'ee d'Italie, 69364 Lyon C\'edex 07, France.\\}

\maketitle

\begin{abstract}
We study the interplay between a $XY$ anisotropy $\gamma$,
exchange modulations and an external magnetic field along the $z$
direction in the $XYZ$ chain using bosonization and Lanczos
diagonalization techniques. We find an Ising critical line in
the space of couplings which occur due to competing relevant
perturbations which are present. More general situations are
also discussed.

\vspace{10 pt}


\vspace{-12 pt}

\end{abstract}

\vskip2pc

\vskip 0.35cm
{\it Introduction}.
The competition between different relevant perturbations
which can render a system critical in a certain domain of the
couplings space has been studied by many authors
\cite{Sierra,nos,Pepino,Shura,Shura2,Shura3,AffleckIT,Totsuka}.
Within a low energy description of many different 1D lattice models,
one often finds the double-frequency sine-Gordon model \cite{Pepino},
that is, a  $U(1)$ scalar field with two perturbations of different
frequencies. In the case where both perturbations are relevant and
for a certain frequency ratio, it has been conjectured that an Ising
criticality can arise \cite{Pepino}. Such a situation has been found
in the study of different lattice systems (see {\it e.g.} \cite{Shura3}
and references therein). More recently, the so-called self-dual
sine-Gordon model has been studied in \cite{Shura3}.

A simple realization of this situation is found in an exactly
solvable case, {\it i.e.} in the dimerized $XY$ chain
\cite{Totsuka,chinos} and a qualitative analysis of the phase
diagram has been presented for more general cases in a field
\cite{Totsuka}. The effect of an $XY$ anisotropy has also been
studied for the Fibonacci $XY$ chain in a magnetic field where it
was shown that the rather involved staircase structure of the
magnetization curve gradually disappears by increasing the
anisotropy of the spin exchange interactions \cite{cochinos}.

In the present paper we present a unified picture of the above
mentioned effects and describe, by using bosonization as well as
numerical techniques, generic situations in which a spin gap
opening mechanism \cite{CGHP} competes with an $XY$ anisotropy to 
render Ising criticality. 
This issue was studied first in \cite{AffleckIT,Totsuka}.
In \cite{Totsuka} the interplay between a gap-opening perturbation
and $XY$ anisotropy was analyzed and a qualitative phase diagram
was proposed. In this paper we focus on the Hamiltonian
\beq
H = \sum_{n} \left((1 + \gamma)S^x_n S^x_{n+1} + (1 - \gamma)
S^y_n S^y_{n+1} + \delta (-1)^n \left(S^x_n S^x_{n+1}+ S^y_n
S^y_{n+1}\right)+\Delta S^z_n S^z_{n+1}\right)-h\sum_n S^z_n\:,
\label{HamXYChinos}
\eeq
which captures most of the essential aspects referred to above,
and provide quantitative results which support the statements
made in \cite{Totsuka}. A more general situation as that arising 
in the Fibonacci $XYZ$ chain in a magnetic field is discussed
in the conclusions.

\vskip 0.35cm
{\it Free Fermion Results.}
In preparation for the analysis of more general situations,
first we discuss the $XY$ or $\Delta = 0$ case which already
bears some generic features. In this case the model is exactly
solvable as it reduces to a bilinear fermionic form.
As is well known, after the usual (1-$d$) Jordan-Wigner
transformation the Hamiltonian can be readily diagonalized by
means of a Bogoliubov transformation, leading to

\beq
H = \sum^{N/4}_{k=1} \left\{ E_{k,0} (\delta,\gamma,h)
\left(d^{\dagger}_{k,0} d_{k,0} + 1\right) + E_{k,1} (\delta,\gamma,h)
\left(d^{\dagger}_{k,1} d_{k,1} + 1\right) \right\}
\eeq
where
\beq
{E_{\!_k}}_{(0,1)} = \frac{1}{\sqrt{2}}\left[1 + \cos(k) + (\delta^2 +
\gamma^2)(1 - \cos(k)) +  2h^2 \pm f(\delta,\gamma,k,h)\right]^{1/2}
\label{bands}
\eeq
and
\beq
f(\delta ,\gamma ,k,h)=2\left[ \delta ^{2}\gamma ^{2}(1-\cos
(k))^{2}+2h^{2}\left( 1+\cos (k)+\delta ^{2}(1-\cos (k))\right)
\right]
^{1/2}.
\label{bands2}
\eeq

The mean value of the magnetization is thus given by

\beq
\langle M \rangle = \frac{\partial \langle H \rangle_{GS}}{\partial h} =
\frac{1}{L}\sum^{N/4}_{k=1} \left[ \frac{\partial E_{k,0}}{\partial h}
(\delta,\gamma,h) + \frac{\partial E_{k,1}}{\partial h}
(\delta,\gamma,h)\right]
\label{mdeh}
\eeq
where GS denotes the ground state, which is the vacuum of Bogoliubov
particles' Fock space. (It should be noticed that the external field
does not act as a chemical potential for these particles).

From Eqs. (\ref{bands}) and (\ref{bands2}) one easily obtains
the critical field values where one of the modes becomes gapless,
{\it i.e} where the Ising transition occurs \cite{Totsuka}

\beq
h_c = \pm \sqrt{\delta^2-\gamma^2} \ ,
\label{critline0}
\eeq
which shows that the Ising transition can only occur for
$\delta > \gamma$.

The exact magnetization curve shown in Fig.\ 1 displays all
the features we want to discuss (we focus in the
region of $h>0$). First of all, one notices that there is no
actual plateau since the magnetization starts to increase as soon
as $h$ is turned on, and this is due to the breaking of the $U(1)$
symmetry for $\gamma \ne 0\,$. In the region of fields below $h_c$
[\,Eq.(\ref{critline0})\,], we are in what we call the ``pseudo
plateau region", where the slope of the magnetization curve is
small and we have dominant density wave correlations. At the
critical field one observes the Ising transition where the
magnetization should behave as $M-M_c \propto
(h-h_c)\left(\log\vert h-h_c\vert - 1 \right)\,$. For $h > h_c$,
the system is in the $XY$ phase where the magnetization increases
more rapidly. Finally, a second Ising transition is observed
before saturation, which occurs at $h=1$ independently of the
values of $\delta$ and $\gamma$. For $\delta < \gamma$ the first
transition does not occur.

\vskip 0.35cm
{\it Bosonization Analysis.}
We discuss the results obtained so far within the bosonization
approach given below, where we argue that the same picture is valid
for arbitrary $\Delta$, provided $\delta$ and $\gamma$ are suitable
renormalized.
For small $\gamma$ and $\delta$ one can study the effects of
such perturbations using bosonization, which allows one to
include $\Delta$ and $h$ exactly through the Bethe Ansatz solution for
the Luttinger parameter $K$.

The large scale behaviour of the $XXZ$ chain can be described by a
$U(1)$ free boson theory with Hamiltonian
\beq
H_0 = {1 \over 2} \int dx \left( v K (\partial_x \tilde\phi)^2
+ {v \over K} (\partial_x \phi)^2 \right) \,,
\label{LL}
\eeq
which corresponds to the Tomonaga-Luttinger Hamiltonian.
The field $\phi$ and its dual $\tilde{\phi}$ are given by the sum
and difference of the light-cone components, respectively. The constant
$K$ governs the conformal dimensions of the bosonic vertex
operators and can be obtained exactly from the Bethe Ansatz
solution of the $XXZ$ chain (see e.g.\ \cite{CHP} for a detailed
summary and references therein). We have $K=1$ for the $SU(2)$
symmetric case ($\Delta = 1$) and is related to the radius
$R$ of \cite{CHP} by $K^{-1} = 2
\pi R^2$. In (\ref{LL}) $v$ corresponds to the Fermi velocity of
the fundamental excitations of the system.
In terms of these fields, the spin operators read

\bea
S_x^z
&\sim& {1 \over \sqrt{2\pi}} \partial_x \phi + a : \cos(2 k_F x +
\sqrt{2 \pi} \phi): + \frac{\langle M \rangle}{2} \, , \label{sz}
\\S_x^{\pm} &\sim& (-1)^x :e^{\pm i\sqrt{2\pi} \tilde{\phi}} \left(b
\cos(2 k_F x + \sqrt{2 \pi} \phi) + c \right) : \, ,
\label{s+}
\eea
where the colons denote normal ordering with respect to the
ground state with  magnetization $\langle M \rangle$. The Fermi
momentum $k_F$ is related to the magnetization of the chain as
$k_F = (1-\langle M \rangle )\pi/2$. The effect of an $XYZ$
anisotropy and/or the external magnetic field is then to modify
the scaling dimensions of the physical fields through $K$ and the
commensurability properties of the spin operators, as can be seen
from  (\ref{sz}), (\ref{s+}). The constants $a$, $b$ and $c$ were
numerically computed in the case of zero magnetic field \cite{HF}
(see also \cite{LA}).

The bosonized Hamiltonian including the perturbations then reads
\beq
H_{bos}=H_0+\lambda_1 \int dx \cos(\sqrt{2\pi}\phi)
+\lambda_2 \int dx
\cos(\sqrt{8\pi}\tilde\phi)
\label{Heff}
\eeq
where $\lambda_1\propto\delta$, $\lambda_2\propto\gamma$.

The scaling dimensions of the perturbations in (\ref{Heff}) are
$K/2$ and $2/K$ respectively, which in the $XX$ case ($K=2$) are
both equal to unity. This allows one to understand within this
approach the appearance of the Ising transition \cite{Shura3}
since in this self-dual case and for $h=0$ one can map the bosonic
system into two Majorana fields, and the critical line is given by
$\lambda_1 = \pm \lambda_2$ where one of the masses of the two
Majorana fields vanishes rendering criticality. As soon as the
external field is turned on the masses of these Majorana fermions
change and we then need a bigger value of
$\lambda_1$ ({\it i.e.} $\delta$) to find the transition \cite{Totsuka}
[see (\ref{critline0})].

This can be extended for $\Delta > 0$, since its only effect is to
modify the scaling dimensions of the two perturbations through
$K(\Delta,h)$. It is interesting to note, however, that for
$\Delta > 0$ the cosine of the dual field is less relevant,
becoming barely marginal for $\Delta = 1$ and $h=0$ ($K(1,0)=1$),
opening the question if one should still expect the Ising
transition to occur. One can argue that this is so by analyzing
the one-loop renormalization group (RG) equations, where it can be
seen that due to the renormalization of $K$, which (up to this
order in the RG) flows to the $XX$ value at large scales 
for certain initial values of the couplings, the competition
between the two cosines can still happen giving rise to the Ising
criticality even in this limiting situation. From this qualitative
analysis one expects a critical line for $h=0$ given by

\beq
\delta \propto \gamma^{\nu}
\label{critline1}
\eeq
with $\nu = \frac{K_{eff}(1-K_{eff}/4)}{(K_{eff}-1)}$ and
$K_{eff}$ is the large scale value of $K$.
As mentioned above, for $0 \le \Delta < 1\,$, the one loop RG
equations indicate that $K_{eff} = 2\,$, therefore
implying that $\nu \equiv 1\,$. A thorough RG analysis should be
performed nevertheless, since this large scale value of $K_{eff}$
is attained in a strong coupling regime, where the analysis ceases
to be valid, but this is out of the scope of the present paper.
We provide numerical evidence for the validity of 
eq.\ (\ref{critline1}) with $\nu = 1$ below.

\vskip 0.35cm
{\it Numerical Results.}
In what follows, we examine this latter conjecture by studying 
numerically the extreme $SU(2)$ case ($\Delta=1$). On the other hand,
this enables an independent test of the bosonization scenario
within non-perturbative regimes.

First, we computed the gap of Hamiltonian (\ref{HamXYChinos})
at $\langle M \rangle = 0\,$ by means of Lanczos diagonalizations
\cite{Lanczos} of finite chain lengths $L$ with periodic boundary
conditions. If $L/4$ is an integer it turns out that the ground
state is even, {\it i.e.} $\exp\,\left( i \pi\, \sum_n \sigma^+_n
\sigma^-_n \right) \equiv 1\,$, otherwise the gap spectrum has to
be computed within the odd subspace of $H$ \cite{Lieb}.
In Fig.\ 2 we display the results so obtained
for $14 \le L \le 22\,$ (see \cite{lengths}\,).
As expected, finite size effects become noticeable on
approaching the critical regime encompassed within the pronounced
gap minima, whereas convergence towards the thermodynamic limit
becomes typically logarithmic \cite {Gutt}.
Interestingly, the locations of these minima result however
fairly independent of the system size, thus facilitating the
estimation of the critical line.  
For $L \le 12\,$ however, the spectrum gap increases monotonically 
with $\vert \gamma \vert$. Therefore, to avoid misleading results 
we dismiss the data of those small sizes. In turn, this impedes
the usage of most of the recursive extrapolation 
algorithms encountered in the literature \cite {Gutt}, 
which show their full strength only if the can be iterated
several times on a large sequence of sizes.
Thus, we content ourselves with a standard (logarithmic) gap 
extrapolation of the form ${\rm g} \simeq {\rm gap}\,(L) \, 
+ \,A /L^{B}\,$. Though semi-quantitative [i.e. 
$A = A(\gamma, \delta)\,$, $B = B (\gamma,\delta)\,$ \,],
the results shown in the inset of Fig.\ 2 nevertheless indicate
a gap closing at the minima of the finite size data.

In studying numerically the massless regime obtained from such minima
however, it should be borne in mind that finite size
corrections to the gap of the homogeneous Heisenberg chain
($\gamma = \delta = 0\,$), vary slowly as $\ln (\ln L)/\ln^2 L\,$,
thus affecting the results over a wide range of sizes
\cite{Griffiths}. Therefore, we are confronted to restricting
considerations to regimes where either $\vert \gamma \vert$ or
$\vert \delta \vert$ are not too small. Fig.\ 3 exhibits our
estimation of the critical line down to $\delta = 0.15\,$ above
which however, a wide linear regime shows up, in agreement with
the bosonization approach. Similar results were observed for
$0 < \Delta <  1\,$ as well.

Secondly, we turn to the GS magnetization curves of Hamiltonian
(\ref{HamXYChinos}). These were calculated numerically as
$\frac{1}{L} \sum_j \langle GS \vert\, \sigma^z_j\,\vert GS \rangle\,$,
where in general the parity $\exp\,\left( i \pi\, \sum_n \sigma^+_n
\sigma^-_n \right)\,$ of $\vert GS \rangle\,$
comes out to be both field and size dependent.
In spite of the existence of a whole massive regime away the critical
line, the magnetization grows linearly for small applied fields $h$
(a feature which holds also for $\Delta = 0\,$),
rather than displaying a plateau at $\langle M \rangle = 0\,$.
This is shown by the inset of Fig.\ 4(a). Since this latter
regime is dominated by a finite gap width (i.e. short range
correlations), finite size effects become negligible there.
This renders our low field results most reliable,
until the first parity change in $\vert GS \rangle\,$ occurs
(signaled by an abrupt increase of the magnetization), probably
associated to the emergence of gapless modes such as those referred
to in the free fermion case.
When selecting the Hamiltonian parameters in a massless (zero field)
point, the magnetization still grows linearly with $h\,$
though with a substantially bigger amplitude, as is shown
by the inset of Fig. 4(b).
This results in the removal of all the pseudo-plateaux observed
in Fig.\ 4(a) with a massive $\gamma-\delta\,$ point.
Also, notice that near the brink of saturation in either case
the susceptibility tends to diverge, alike the $XY$ situation.

Finally, we address to the predicted Ising behavior of the
magnetization curves near the critical fields $h_c\,$ discussed
in the previous sections.
Upon estimating the former with the fields $h_c (L)$ involving the
first parity jump in the GS of a finite chain,
we obtained a fair logarithmic regime actually applying over more than
two decades in $(h-h_c)\left(\log \vert h-h_c \vert - 1 \right)\,$.
This is displayed in Fig.\ 5. The value of the upper line slope
results almost independent of the system size and is indicative of
the logarithmic regimes entailed both by the free fermion and
bosonization approaches.

\vskip 0.35cm
{\it Conclusions.} To summarize, we have studied how the tendencies
towards the formation of massive spin excitations (through dimerized
$\delta$-exchanges), and towards the $XY$-ordering
(via pairing of $\gamma$-interactions), compete with each other.
In the bosonization picture, these tendencies manifest themselves
in the existence of two (competing) relevant interactions and bring
about the Ising transition, which in fact is connected
to the same one driven by an external field.
Due to the breaking of the $U(1)$ symmetry, the latter
does not couple to any conserved quantity and therefore
the magnetization process gets essentially modified with respect
to the $\gamma = 0\,$ case, i.e. it ever increases in spite
of the presence of massive regimes.

Both the bosonization and the numerical analyses support the picture 
that the physical mechanism rendering the Ising transition 
is quite general and valid both in the strong and weak coupling limits. 
Following the ideas given in \cite{nos2} 
one can conjecture that a similar picture for each plateaux in the
XYZ Fibonacci chain holds. The mechanism in that case is the same, 
where the relevant operator coming from dimerization is replaced 
for each plateaux by the operator {\it commensurate} at the 
corresponding frequency of the Fibonacci Fourier spectrum.
From this one concludes that a similar picture as that found 
in the XY case \cite{cochinos} is also valid for generic XYZ 
Fibonacci chains.

The numerical analysis lent further support to the bosonization
results within several non-perturbative situations. 
We found evidence of a massless line along with the expected
Ising like behaviour near critical fields. In turn, this suggests
that most basic features of the fully interacting system can be
captured by the free fermion picture discussed above. 
However, the issue as to whether or not
{\it all} of the GS parity changes induced by the field
[Fig. 4(a)\,] correspond to Ising transitions in the
thermodynamic limit, remains quite open.

\vskip 0.35cm
\noindent {\it Acknowledgments.} It is a pleasure to
acknowledge fruitful discussions with A.\ Honecker and A.A.\ Nersesyan.
The authors acknowledge partial financial support of CONICET
and Fundaci\'on Antorchas.

\vskip 0.35cm

\noindent\small{$\#$ On leave from Universidad
Nacional de La Plata and Universidad Nacional de Lomas de Zamora.}


\begin{twocolumn}

\newpage

\begin{figure}
\hbox{ \epsfxsize=3.5in \epsffile{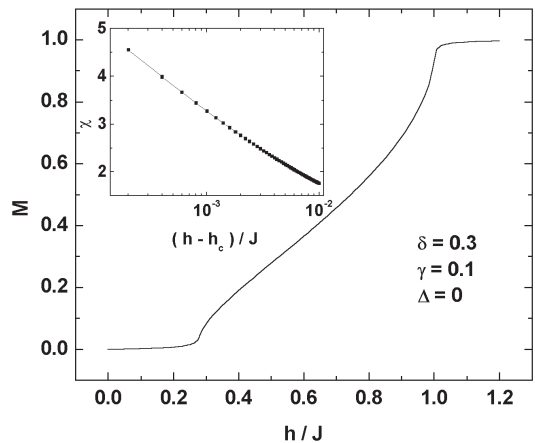}} \vspace{1cm}
\caption{Exact magnetization curve of the $XY$ case. The
inset displays the logarithmic singularity dominating
the susceptibility at $h_c = \sqrt{\delta^2-\gamma^2} \,$.}
\end{figure}
\vspace{.5cm}

\begin{figure}
\hbox{ \epsfxsize=3.5in \epsffile{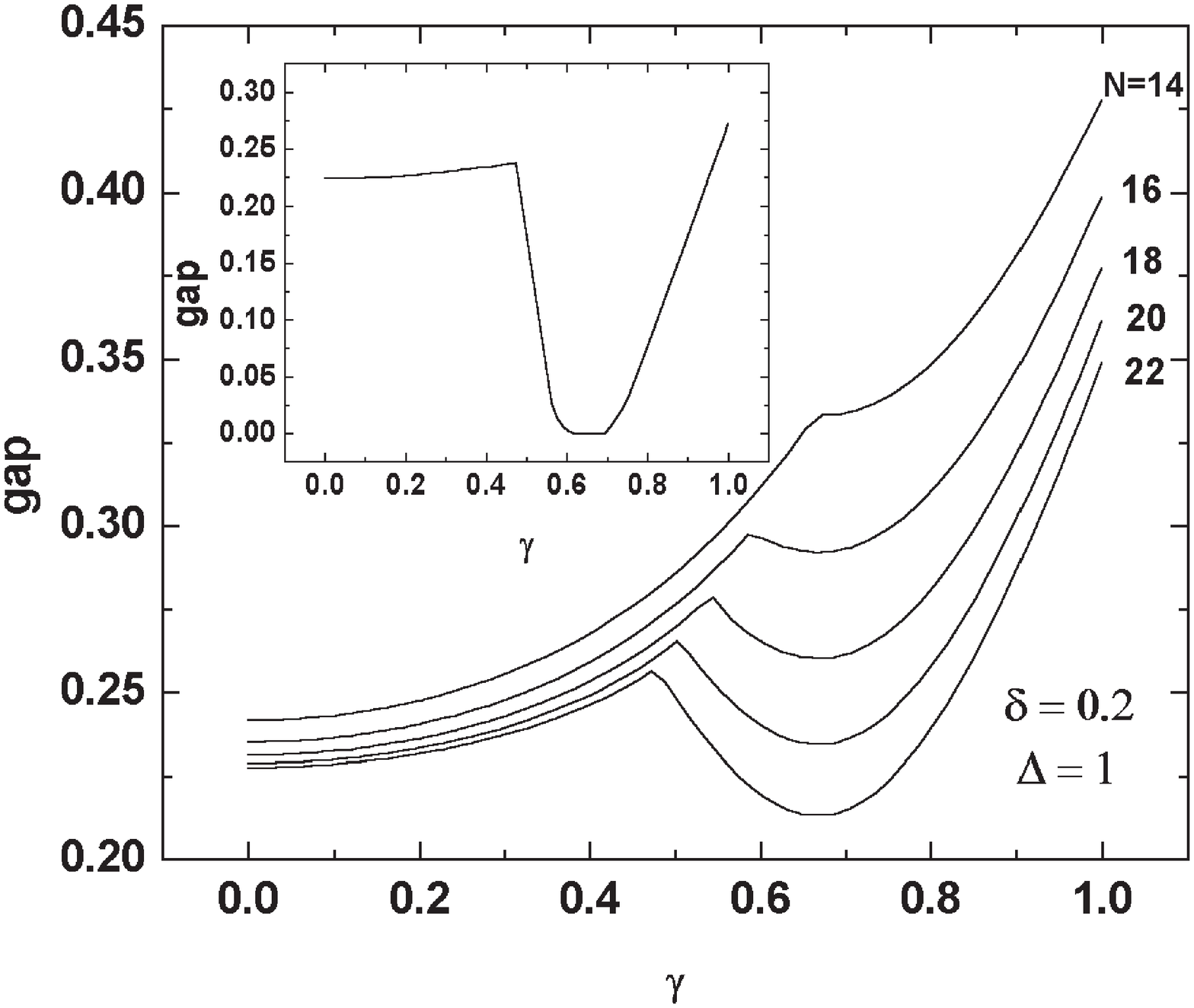}} \vspace{1cm}
\caption{Gap spectrum for different sizes of Hamiltonian (1) with
$\Delta = 1\,$. The inset exhibits the (estimative)
gap extrapolations to the thermodynamic limit.}
\end{figure}

\begin{figure}
\hbox{ \epsfxsize=3.5in \epsffile{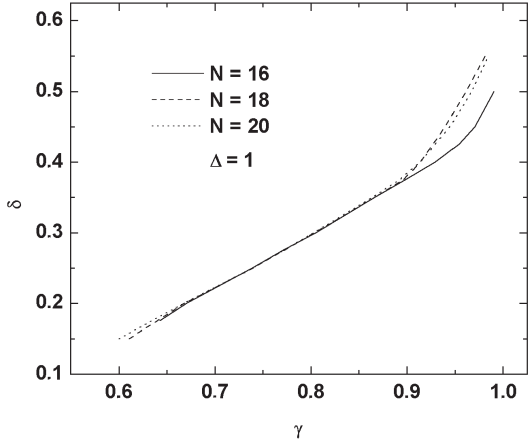}} \vspace{1cm}
\caption{Gapless line of Hamiltonian (1) in the ($\gamma,
\delta\,$) coupling parameter space arising from
the positions of the gap minima of finite samples
with $\Delta = 1\,$ and $\langle M \rangle = 0\,$.}
\end{figure}

\begin{figure}
\hbox{ \epsfxsize=3.5in \epsffile{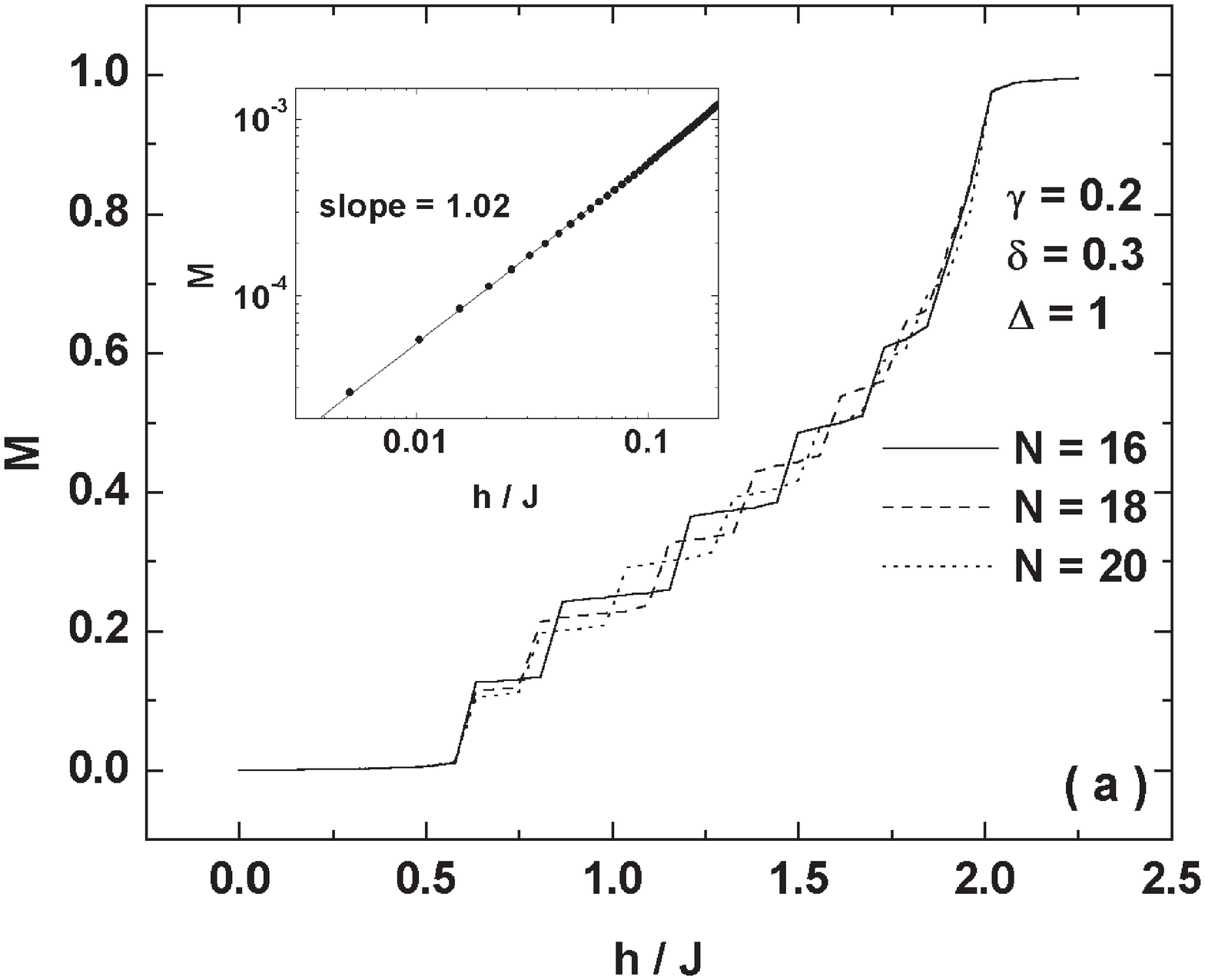}}
\hbox{ \epsfxsize=3.5in \epsffile{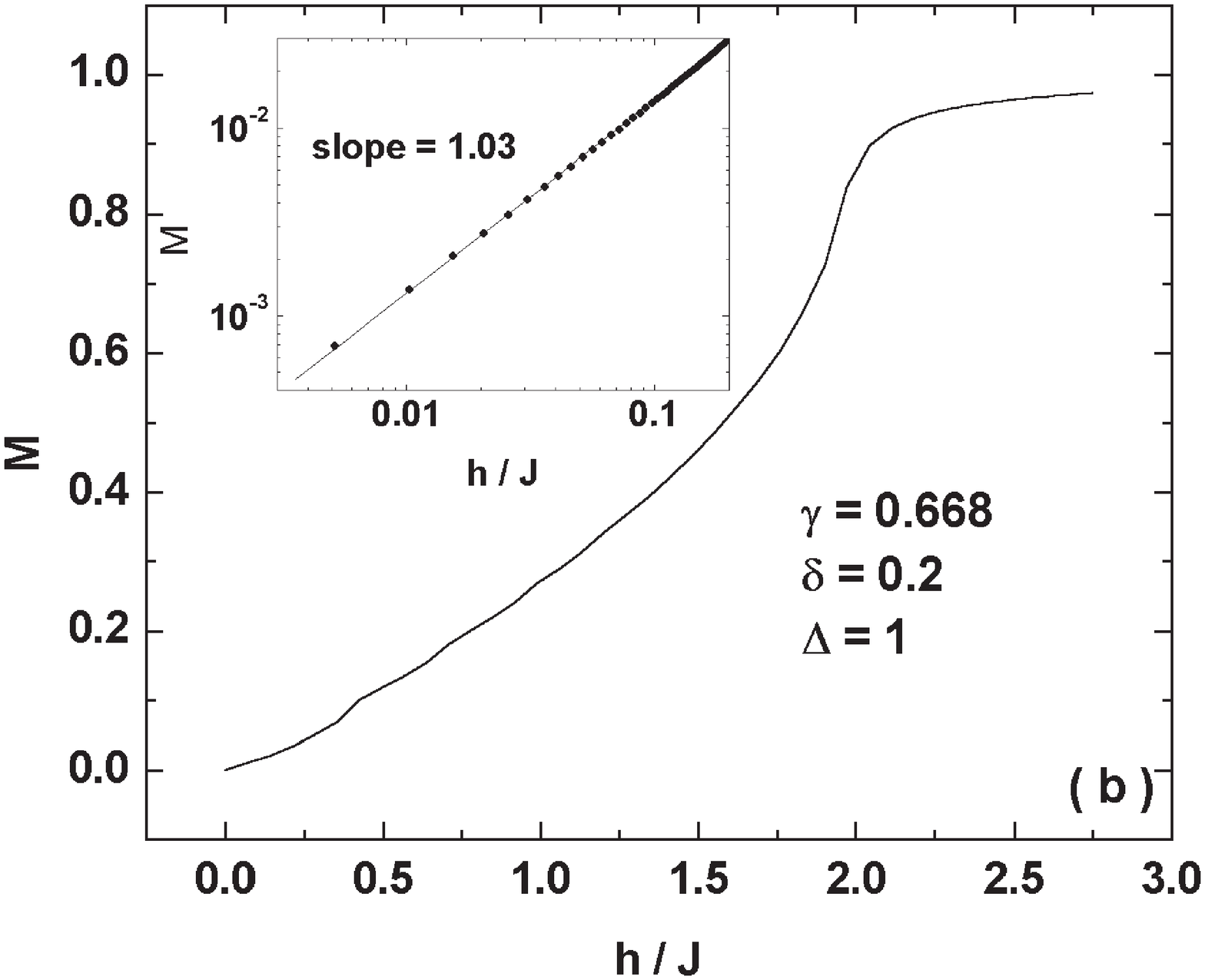}}
\vspace{1cm}
\caption{Magnetization curves for $\Delta = 1\,$. (a) The piece-wise
continuous behavior is related to parity changes in the ground
state. (b) Results arising for a massless
($\gamma, \delta\,$) point. In each case, the inset exhibits
the linear response at low-field regimes.}
\end{figure}

\begin{figure}
\hbox{ \epsfxsize=3.5in \epsffile{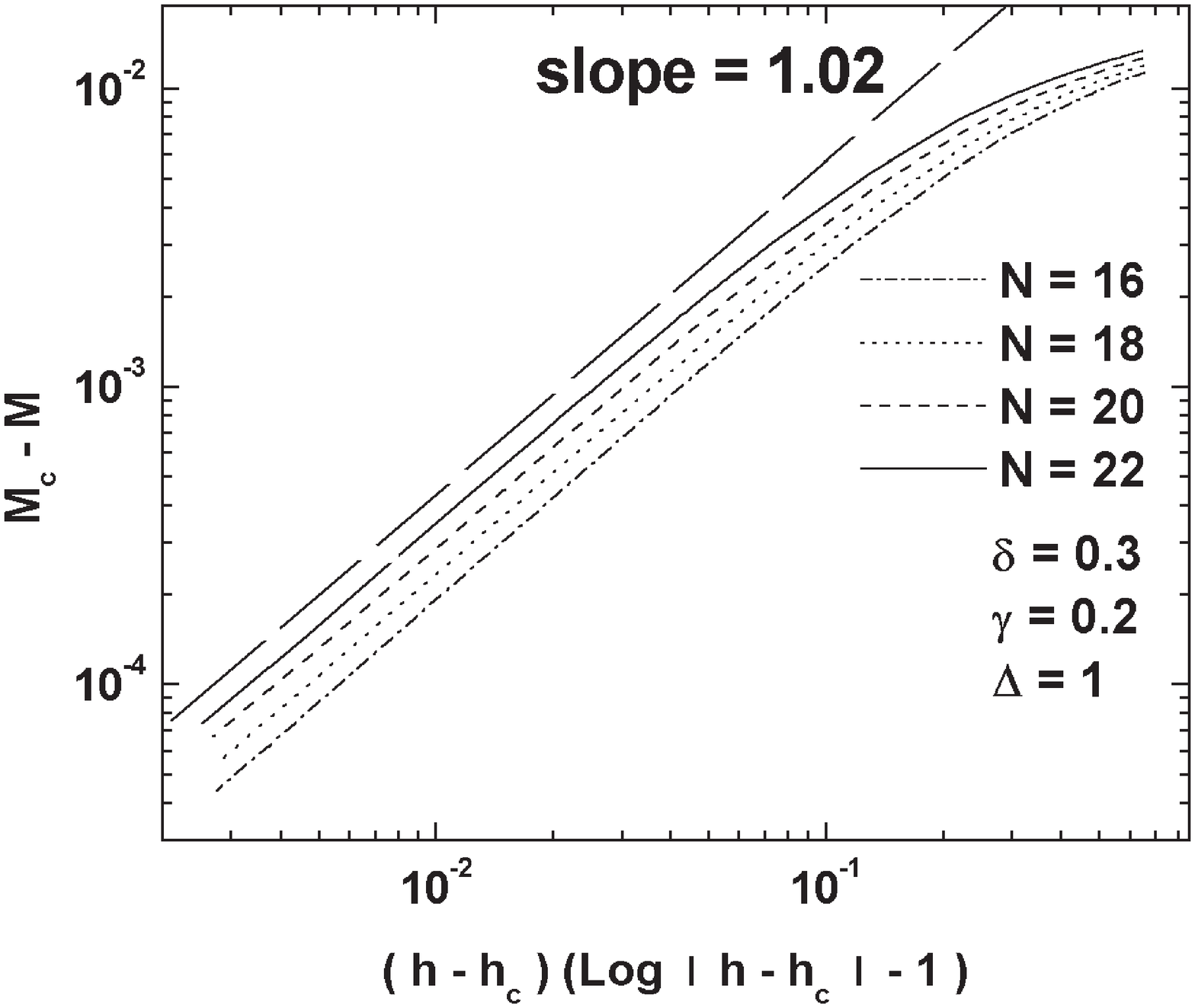}} \vspace{1cm}
\caption{Magnetization behavior for $\Delta = 1\,$ near
critical fields referred to in the text. The slope value
suggests the emergence of a logarithmic regime as that
observed in Fig. 1.}

\end{figure}
\end{twocolumn}


\begin{thebibliography}{99}
\bibitem{lengths} The lengths within our computational reach
are in part limited by the breaking of the $U(1)\,$ symmetry
for $\gamma \ne 0\,$.
\end{thebibliography}
\end{document}